\RequirePackage{fix-cm}
\documentclass[smallextended]{svjour3}       
\smartqed  
\usepackage{amssymb}
\usepackage{amsmath}
\usepackage{comment}
\usepackage{captcont} 
\usepackage{graphicx}
\usepackage{caption}
\usepackage{subcaption}
\usepackage{array}
\usepackage{epstopdf}
\usepackage{algorithm}
\usepackage[noend]{algpseudocode}

\makeatletter
\def\BState{\State\hskip-\ALG@thistlm}
\makeatother
\begin{document}

\title{Performance Tuning of a Parallel 3-D FFT Package OpenFFT 
}

\titlerunning{Performance Tuning of a Parallel 3-D FFT Package OpenFFT}        

\author{Truong Vinh Truong Duy \and Taisuke Ozaki 
}


\institute{
              Institute for Solid State Physics, The University of Tokyo, Kashiwanoha 5-1-5, Kashiwa, Chiba 277-8581, Japan \\
              Tel.: +81 4 7136 3279\\
              \email{duytvt@issp.u-tokyo.ac.jp}           
}

\date{Received: date / Accepted: date}

\maketitle

\begin{abstract}
The fast Fourier transform (FFT) is a primitive kernel in numerous fields of science and engineering. OpenFFT is an open-source parallel package for 3-D FFTs, built on a communication-optimal domain decomposition method for achieving minimal volume of communication. 
In this paper, we analyze and tune the performance of OpenFFT, paying a particular attention to tuning of communication that dominates the run time of large-scale calculations. We first analyze its performance on different machines for an understanding of the behaviors of the package and machines. Based on the performance analysis, we develop six communication methods for performing communication with the aim of covering varied calculation scales on a variety of computational platforms. OpenFFT is then augmented with an auto-tuning of communication to select the best method in run time depending on their performance. Numerical results demonstrate that the optimized OpenFFT is able to deliver relatively good performance in comparison with other state-of-the-art packages at different computational scales on a number of parallel machines. 


\end{abstract}

\section{Introduction}
\label{Introduction}

The fast Fourier transform (FFT) \cite{citeulike:1436613} is an essential primitive for numerous fields of science and engineering, such as electronic structure calculations, digital signal processing, medical image processing, communications, astronomy, geology, and optics \cite{Haynes2000130,Clarke199214,Broughton2008,Gonzales1992}. In these applications, the FFT is usually performed with a large data set in multiple dimensions for multiple times, for instance in 3 dimensions with 3-D FFTs, making it a computationally expensive calculation. Given the importance of the FFT and widespread of massively parallel computers with multi-core processors, there have been a lot of efforts to parallelize the FFT, where the multi-dimensional FFT data is delivered by a decomposition method to the processes so that they can have enough data to perform the 1-D FFTs locally in one specific dimension in parallel. The focus of this paper is the parallelization of 3-D FFTs. State-of-the-art parallel FFT packages for 3-D FFTs range from the 1-D decomposition (FFTW MPI Version \cite{FFTW} and Intel MKL Cluster \cite{MKL}) to the 2-D decomposition (FFTE \cite{takahashi2010implementation,ffte}, P3DFFT \cite{doi:10.1137/11082748X,p3dfft}, PFFT \cite{doi:10.1137/120885887,pfft} and 2DECOMP\&FFT \cite{Li2010,2DECOMP}). 

The 1-D decomposition \cite{Haynes2000130,Dmitruk20011921}, divides the 3-D data into blocks of equal numbers of complete $ab$-planes, for example, along the $c$-dimension to allocate to the processes. The alphabet hereafter is used to denote the dimensions, i.e., $a$ for the first dimension, $b$ for the second dimension, and so on. The 1-D decomposition requires only one transpose step with minimal volume of communication, but the applicable number of processes is limited to the size of one dimension. 
The 2-D decomposition \cite{Ayala2012,takahashi2010implementation} evenly divides the $ab$-plane to the processes, with each having all the data along the remaining $c$-dimension, and thus, offering higher scalability than the 1-D decomposition, as the restriction on the applicable number of processes is now lifted to make it up to the size of two dimensions, but with the cost of incurring higher volume of communication. 



For large-scale 3-D FFT calculations, communication dominates the run time of applications. Since a smaller volume of communication is desired for a better performance, the volume of communication is one of crucial factors in the decomposition method, which requires communication for undertaking the data transpose repeatedly until the data of all the dimensions has been FFT-transformed. Yet on this front, the existing methods incur a considerable volume of communication, caused by small data locality and pre-defined degree of decomposition. The data locality is small because the dimensions involved in the decomposition are treated in the same way, and consequently the order of transpose does not have any impact on the volume of communication, resulting in a relatively small amount of data localized when transposing. Also, the degree of the decomposition is usually pre-defined, regardless of the number of processes, in particular, the 2-D decomposition partitions in two dimensions, even when the number of processes is smaller than the size of one dimension. In fact, as touched on above, the lower the degree of decomposition is, the smaller the volume of communication is. Hence, the present methods could not take advantage of a lower degree decomposition. A communication-aware decomposition method should localize as much data as possible, and be adaptive to switch between lower and higher degrees of decomposition depending on the number of processes to reduce the volume of communication.

To address the problem with the volume of communication, we have developed a communication-optimal decomposition method for the parallelization of multi-dimensional FFTs, achieving the smallest volumes of communication for all ranges of the number of processes compared to the currently used methods by two distinguished features: adaptive decomposition and transpose order awareness \cite{Duy2014153}. In our method, the FFT data is decomposed based on a row-wise basis that translates the corresponding coordinates from  multi-dimensions into one-dimension so that the one-dimensional data can be divided and allocated equally to the processes using a block distribution for a good load balance among them. As a result, the method can adaptively decompose the FFT data on the lowest possible degree according to the number of processes, and thus, reducing the volume of communication in the first place. Furthermore, as we treat the dimensions engaged in the decomposition differently, different orders of transpose actually incur different degrees of data locality. The best transpose orders that can localize large amounts of data when transposing leading to the smallest volumes of communication for the 3-D, 4-D, and 5-D FFTs are identified by analyzing all possible cases. 

Based on the method, we have developed and released a parallel package for 3-D FFTs, called OpenFFT \cite{openfft,duytvtisc14}, in C and MPI with support for Fortran through the Fortran interface. 
Numerical results have shown its good performance and scaling property \cite{Duy2014153}. As a practical application example, OpenFFT has been used in a density functional theory code for nano-scale materials simulations called OpenMX \cite{Duy2014777,openmx}.

Although the performance evaluation of OpenFFT has been conducted in \cite{Duy2014153}, performance tuning and comparison of OpenFFT are definitely further required by the following reasons. First, although having achieved minimal volume of communication is an advantage, development of efficient communication methods is undeniably of great importance. In \cite{Duy2014153}, which is equivalent to OpenFFT version 1.0, we developed and implemented only one communication method by posting the non-blocking pairs of MPI\_Isend() and MPI\_Irecv(), followed by MPI\_Waitall() to wait for all processes. This communication method is not expected to be well suited to any calculation scale on any machine, and tuning of communication by developing more flexible and efficient communication methods is necessary as a consequence. Second, the evaluation in \cite{Duy2014153} was only carried out on a single computational platform, particularly the Cray XC30 machine. As machine specifications, namely the CPU, interconnection network, compiler, MPI library, memory access speed, and cache miss penalty of machines differ from one to another, performance analysis, tuning, and evaluation must be extended to a wider variety of machines. Finally, the performance comparison in \cite{Duy2014153} was also thought to be incomplete, because even though it contained three third-party packages, there was only one package adopting the 2-D decomposition. Hence, more packages with the 2-D decomposition have to be included for a more rounded comparison. Such thorough performance evaluation and comparison are expected to serve as a useful reference for potential users who are looking for a high performance parallel 3-D FFT library to employ in their applications. 

In this paper, we conduct performance analysis and tuning of OpenFFT version 1.0, with a primary focus on the development of communication methods, and then comprehensively compare the optimized OpenFFT version 1.1 with three well-known 2-D decomposition packages for 3-D FFTs (FFTE \cite{takahashi2010implementation,ffte}, P3DFFT \cite{doi:10.1137/11082748X,p3dfft}, and 2DECOMP\&FFT \cite{Li2010,2DECOMP}) on four different machines (the Cray XC30, SGI InfiniBand, Fujitsu FX10, and K computer) across a range of computational scales. We first analyze the performance of OpenFFT on the machines for understanding the behaviors of the package and machines. Given the performance analysis, we develop six communication methods in OpenFFT, ranging from all-to-all to point-to-point communication routines, overlapping and non-overlapping communication with computation, and grouping the processes and communication. We then compare the performance of the communication methods on the machines with different interconnects, and augment OpenFFT with an auto-tuning of communication for dynamically choosing the best method. Additionally, allocation and deallocation of arrays are also optimized. Numerical results demonstrate that OpenFFT version 1.1 can achieve relatively good performance with different machines and computational scales. The communication methods are actually general, and can be implemented and applied in parallel applications.

The remainder of the paper is organized as follows. Section 2 briefly introduces the domain decomposition method of OpenFFT. Performance analysis and tuning are presented in Section 3. Section 4 gives numerical comparison in terms of the volume of communication and time-to-solution between OpenFFT and other packages. Finally, our study is concluded in Section 5.

\section{Domain decomposition method of OpenFFT}
\label{Methods}

For the sake of clarity and for making this paper relatively self-contained, in this section we recall the domain decomposition method used in OpenFFT. A full description of the method can be found in \cite{Duy2014153}. 

Assuming that the numbers of data points along the $a$-, $b$-, and $c$-axes are $N_{\mathrm{1}}$, $N_{\mathrm{2}}$, and $N_{\mathrm{3}}$, respectively, the number of processes is $N_p$, and $myid$ is the process identification defined to be in the range of [0, $N_p-1$]. Our method involves a number of steps as follows. 

\begin{figure}
        \centering
        \begin{subfigure}{\textwidth}
                \includegraphics[width=\textwidth]{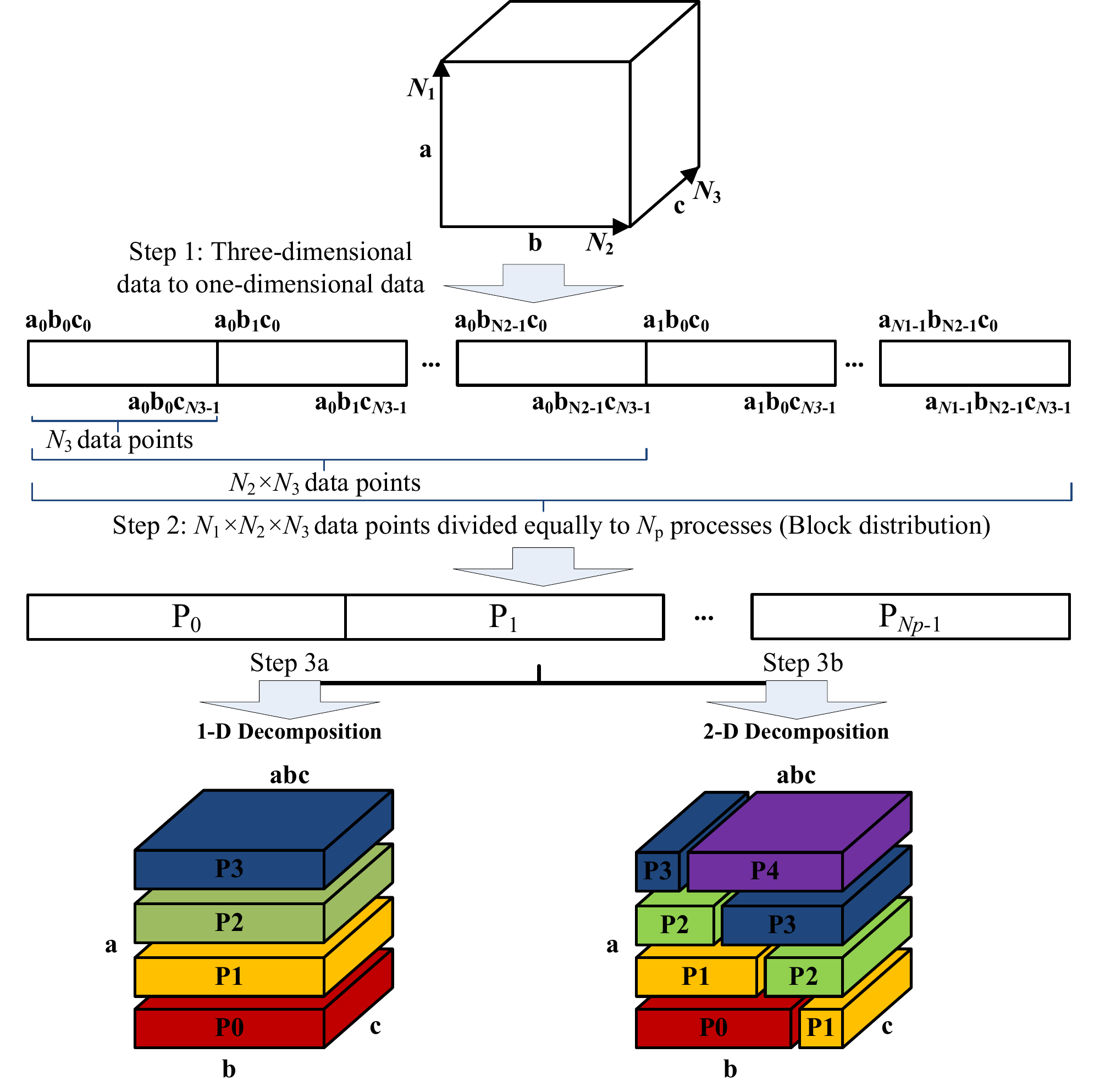}
                \caption{Adaptive decomposition: example of the $abc$ decomposition.}
                \label{fig-3D-1D}
        \end{subfigure}%
        
        \begin{subfigure}{\textwidth}
                \includegraphics[width=\textwidth]{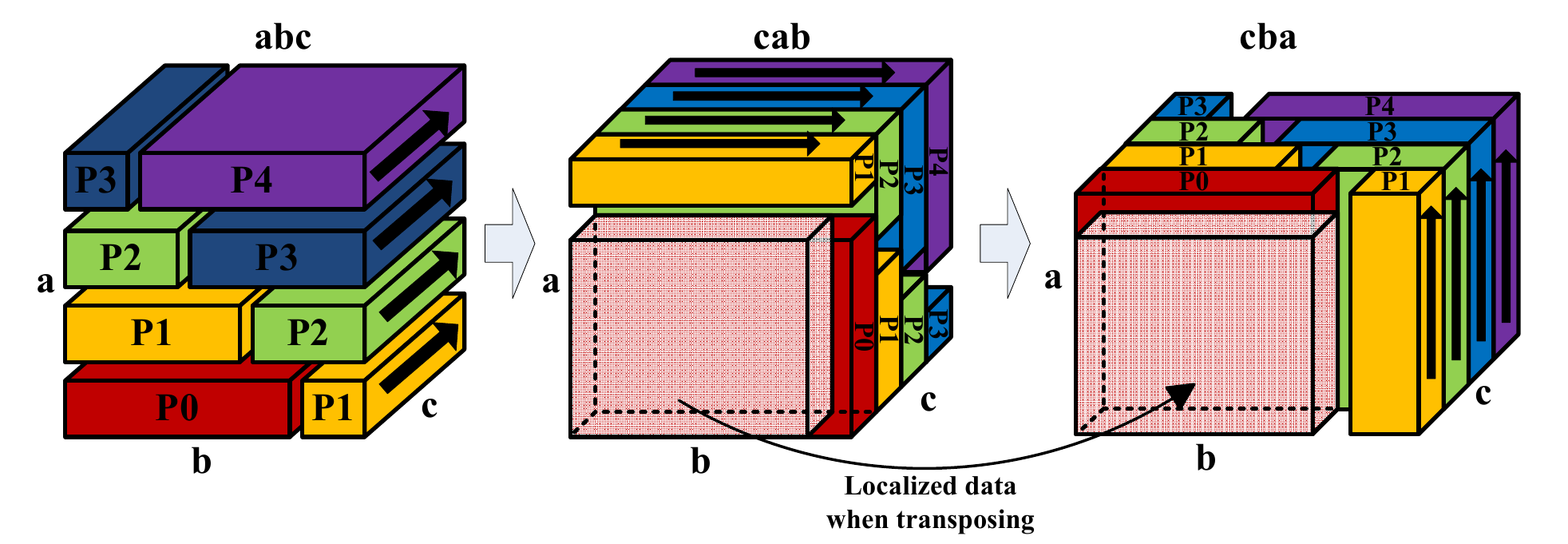}
                \caption{Transpose-order awareness: transpose from $cab$ to $cba$, a large volume of data can be localized.}
                \label{fig-3D-2Da}
        \end{subfigure}
        \caption{Domain decomposition method of OpenFFT.}\label{fig-domain-openfft}
\end{figure}

First, we translate the original 3-D FFT data $A(a,b,c)$ into the 1-D data $X(x)$ (Step 1 in Fig. \ref{fig-3D-1D}, with the $abc$ decomposition as an example). 
The relationship between a 3-D coordinate $A(a_1,b_1,c_1)$ and a 1-D coordinate $X(x_1)$ in the $abc$ decomposition is given by
\begin{equation}
x_1=a_1 \times N_2 \times N_3 + b_1 \times N_3 + c_1.
\end{equation}

A function $f()$ for translating the 3-D and 1-D coordinates to identify the starting and ending points allocated to each process is defined as    
\begin{equation}
f(N_1,N_2,N_3,N_p,myid) = \begin{cases} \left \lfloor \frac{N_1 \times myid}{N_p} \right \rfloor \times N_2N_3, & \mbox{if } N_p \leq N_1 ;
\\ 
\\ \left \lfloor \frac{N_1N_2 \times myid}{N_p} \right \rfloor \times N_3 , & \mbox{if } N_1 < N_p \leq N_1N_2 ,
\end{cases}
\end{equation}
where $\left \lfloor  \right \rfloor$ is the floor function. 

We then equally divide the 1-D data to the processes using a block distribution (Step 2 in Fig. \ref{fig-3D-1D}), in which a process with $myid$ is assigned the data points from $X(x^{s}_{myid})$ to $X(x^{e}_{myid})$ in one dimension, where
\begin{equation}
x^{s}_{myid}=f(N_1,N_2,N_3,N_p,myid),
\end{equation}
\begin{equation}
x^{e}_{myid}=f(N_1,N_2,N_3,N_p,myid+1)-1.
\end{equation}

These 1-D coordinates can be translated back to the 3-D ones to obtain the corresponding starting and ending coordinates in three dimensions:  
\begin{equation}
\label{eq-3D-ase}
a^{(s,e)}_{myid}=\left \lfloor \frac{x^{(s,e)}_{myid}}{N_2N_3} \right \rfloor ,
\end{equation}
\begin{equation}
\label{eq-3D-bse}
b^{(s,e)}_{myid}=\left \lfloor \frac{x^{(s,e)}_{myid}-a^{(s,e)}_{myid}N_2N_3}{N_3} \right \rfloor ,
\end{equation}
\begin{equation}
\label{eq-3D-cse}
c^{(s,e)}_{myid}=x^{(s,e)}_{myid}-a^{(s,e)}_{myid}N_2N_3-b^{(s,e)}_{myid}N_3,
\end{equation}
where $A(a^{s}_{myid},b^{s}_{myid},c^{s}_{myid})$ and $A(a^{e}_{myid},b^{e}_{myid},c^{e}_{myid})$ are the starting and ending points, respectively, in three dimensions for a process with $myid$.

Consequently, the decomposition has two forms depending on the number of processes. The distribution of the data points is carried out in either 1-D (Step 3a in Fig. \ref{fig-3D-1D}) or 2-D (Step 3b in Fig. \ref{fig-3D-1D}), determined by the first one or two dimensions, respectively. For instance, with $N_p$ processes and $N_1 < N_p \leq N_1N_2$, the $abc$ decomposition takes place in 2-D, where a process with $myid$ is allocated the data points from $A(a^{s}_{myid},b^{s}_{myid}, 0)$ to $A(a^{e}_{myid},b^{e}_{myid}, N_{\mathrm{3}}-1)$ in ascending order of the $a$-, $b$-, and $c$-coordinates, where $a^{s}_{myid}$, $b^{s}_{myid}$, $a^{e}_{myid}$, and $b^{e}_{myid}$ can be obtained from Eqs. (\ref{eq-3D-ase}) and (\ref{eq-3D-bse}), with
\begin{equation}
x^{s}_{myid}=\left \lfloor \frac{N_1N_2 \times myid}{N_p} \right \rfloor \times N_3,
\end{equation}
\begin{equation}
x^{e}_{myid}=\left \lfloor \frac{N_1N_2 \times (myid+1)}{N_p} \right \rfloor \times N_3-1.
\end{equation}

Next, we have to follow a good transpose order to localize as much data as possible when transposing. Figure \ref{fig-3D-2Da} exemplifies such a good order, $abc \rightarrow cab \rightarrow cba$, which is currently utilized in OpenFFT. The figure shows a large overlap between the areas distributed to, for example process P0, in the $cab$ and $cba$ decompositions, indicating that a large amount of data is already localized and can be reused when transposing from $cab$ to $cba$, leaving just a small amount of data that needs to be communicated with other processes.


\section{Performance Analysis and Tuning}

\subsection{Performance Analysis}

\subsubsection{Calculation Flow}

\begin{figure}[htbp]
\begin{center}
\includegraphics[width=1.0\textwidth]{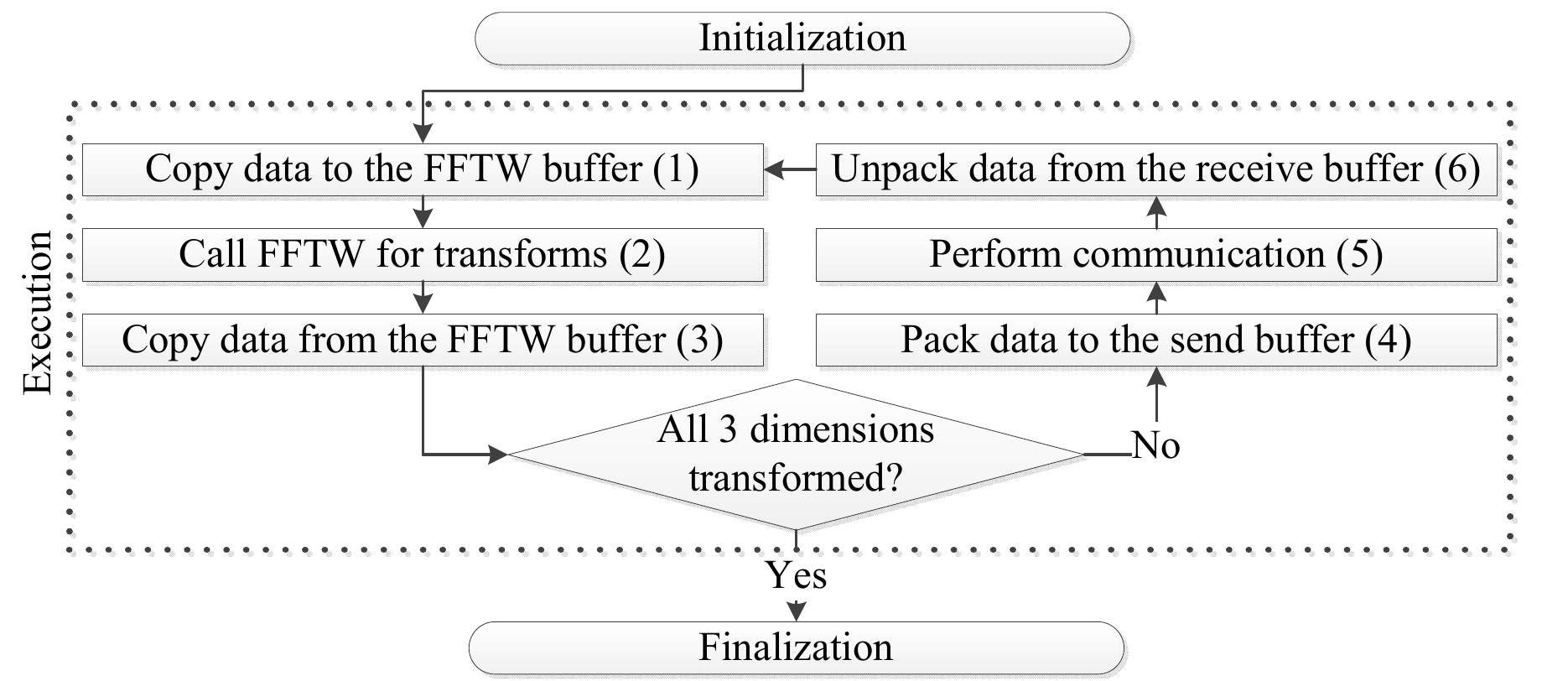}
\end{center}
\caption{Calculation flow of OpenFFT.}
\label{fig-cal-flow}
\end{figure}

Figure \ref{fig-cal-flow} shows the calculation flow of OpenFFT that is comprised of three main phases: initialization, execution, and finalization. In the initialization phase, important variables are initialized, including the number of data points and the global indexes of the data points allocated to a process upon starting, the number of data points and the global indexes of the data points allocated to a process upon finishing, and some other control parameters. They are used for allocating and initializing local input data arrays from the global input array, and for allocating and obtaining local output data arrays to gather the global output array. It should be noted that the auto-tuning of communication in OpenFFT version 1.1 is implemented in this initialization phase.  Next, the execution phase, which is the main routine, is taken to perform the transformation. It can be undertaken as many times as necessary. Lastly, the calculation is finalized in the finalization phase, where the memory is freed, and the parameters are reset. Generally, the initialization and finalization phases are excluded from the total elapsed time, as the execution phase is usually performed for multiple times and becomes a dominant factor in practice. 

As can be seen from the figure, the execution itself consists of several steps in each process. First, the local input data is copied to the FFTW buffer, and then the sequential FFTW routine is called for transforms in a particular dimension. The transformed data is then copied back to the output data array from the FFTW buffer. If all the three dimensions have been transformed, the calculation will be completed. Otherwise, if there is any remaining dimension, a transpose step will be conducted for another round of transform in that dimension. The data that is required by and needs to be sent to other processes is packed to the send buffer, and the receive buffer for storing the data received from other processes is also constructed. Then, communication is performed for sending and receiving data among the processes. After that, the data from the receive buffer is unpacked and combined with the data in the local input data for transforming.  

\subsubsection{Breakdown of Execution Time}

Figure \ref{fig-cal-flow} indicates that there are several operations involving computation and communication during a calculation, and hence, the execution time must be broken down to reveal their contribution for understanding their behavior. Furthermore, as machine specifications differ from one to another, the performance must be analyzed on a variety of machines. Table \ref{tab-spec-machine} details the specifications of the machines in use.

\begin{table}
\begin{center}
\caption{Machine specifications.}
\label{tab-spec-machine}
\begin{tabular}{| c || m{0.15\textwidth} | m{0.15\textwidth} | m{0.15\textwidth} | m{0.15\textwidth} |}
\hline
Specification & Cray XC30 & SGI InfiniBand & Fujitsu FX10 & K   \quad\quad\quad Computer \\
\hline
CPU           & Intel Xeon E5-2670 2.6GHz 8 cores $\times$ 2 & Intel Xeon E5-2680v2 2.8GHz 10 cores $\times$ 2 &  Fujitsu SPARC64 IXfx 1.848GHz 16 cores  & Fujitsu SPARC64 VIIIfx 2.0GHz 8 cores   \\
\hline
Memory        &     64GB  &   64GB   & 32GB &   16GB        \\
\hline
Interconnect  & Dragonfly (8.5GB/s) &  InfiniBand 4X FDR (6.8GB/s) &  Tofu (5.0GB/s)   &   Tofu (5.0GB/s)    \\
\hline
Compiler     &     Intel &  Intel   &  Fujitsu  &     Fujitsu       \\
\hline
FFTW          &  3.3.0.4 &  Wrappers by the MKL &   3.3   &   3.3      \\
\hline
\end{tabular}
\end{center}
\end{table}

\begin{figure}[htbp]
        \begin{subfigure}{\textwidth}
                \centering
                \includegraphics[width=0.75\textwidth]{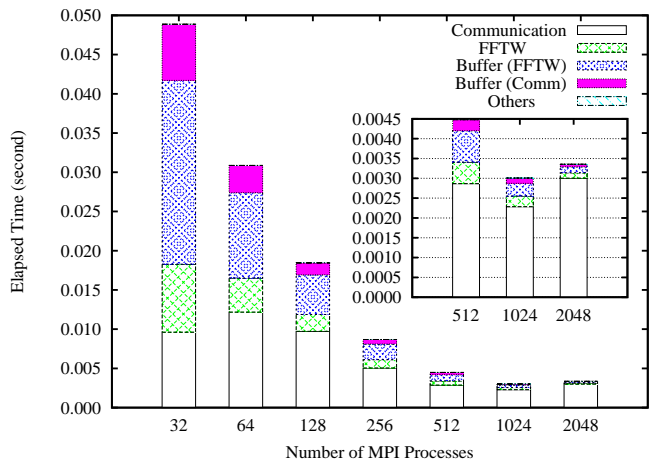}
                \caption{Cray XC30. The inset enlarges the right part of the main graph.}
                \label{fig-breakdown-cray}
        \end{subfigure}%
        
        \begin{subfigure}{\textwidth}
                \centering
                \includegraphics[width=0.75\textwidth]{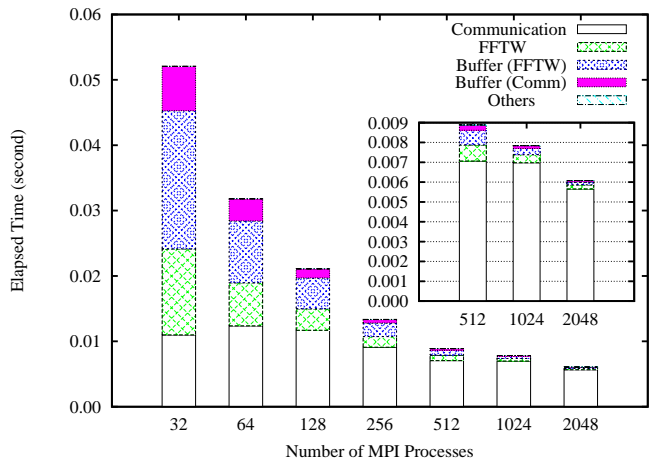}
                \caption{SGI InfiniBand. The inset enlarges the right part of the main graph.}
                \label{fig-breakdown-hster}
        \end{subfigure}

        \begin{subfigure}{\textwidth}
                \centering
                \includegraphics[width=0.75\textwidth]{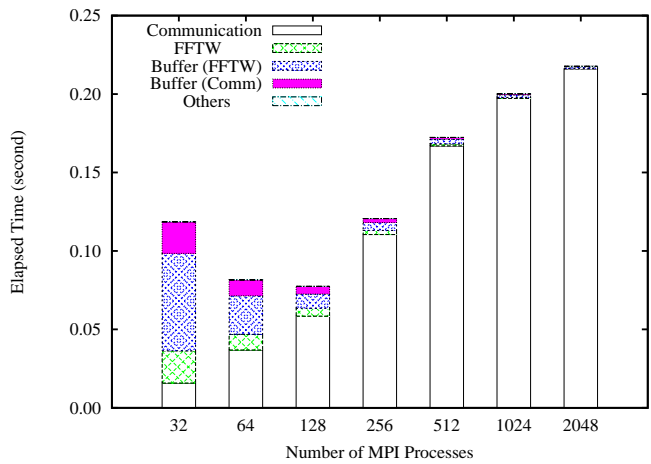}
                \caption{Fujitsu FX10.}
                \label{fig-breakdown-fx10}
        \end{subfigure}%
        \caption{Breakdown of OpenFFT version 1.0 execution time with $256^3$ data points on the Cray XC30, SGI InfiniBand, and Fujitsu FX10.}\label{fig-breakdown-time}
\end{figure}

Figure \ref{fig-breakdown-time} displays the breakdown of the total execution time of OpenFFT version 1.0 with double-precision complex-to-complex transforms of $256^3$ data points on the Cray XC30 (Fig. \ref{fig-breakdown-cray}), SGI InfiniBand (Fig. \ref{fig-breakdown-hster}), and Fujitsu FX10 (Fig. \ref{fig-breakdown-fx10}). Since the specifications of the K computer are basically the same as those of the FX10 and partly owing to resource constraints, the K computer is preserved for large scale evaluations later. The execution phase is performed for ten times, and the summation of the ten longest process times for each operation is averaged and reported. The size of $256^3$ is chosen as a representative case that can divulge general behavior of OpenFFT on the machines. The operations displayed in the figure are described below. 

\begin{itemize}
\item
\textbf{Communication}: the time for performing communication (step (5) in Fig. \ref{fig-cal-flow}). In OpenFFT version 1.0, communication is implemented through the use of MPI\_Isend() and MPI\_Irecv() in combination with MPI\_Waitall(). 
\item
\textbf{FFTW}: the time for performing the 1-D FFTs with FFTW (step (2) in Fig. \ref{fig-cal-flow}). 
\item
\textbf{Buffer (Comm)}: the time for packing data to the send buffer and for unpacking data from the receive buffer (steps (4)+(6) in Fig. \ref{fig-cal-flow}).  
\item
\textbf{Buffer (FFTW)}: the time for copying data to and from the FFTW buffer (steps (1)+(3) in Fig. \ref{fig-cal-flow}).  
\item
\textbf{Others}: other times rather than the above in the execution time.
\end{itemize}

With the exception of the communication time, other times scale almost perfectly to the number of processes. This is just as expected since they are totally pure computations, and the data is equally assigned to the processes by the decomposition method. The problem here is obviously communication that is extremely inefficient with the Tofu interconnect of the FX10, where its time keeps increasing to the number of processes, although it is reasonably satisfactory on the Dragonfly interconnect of the XC30 and the InfiniBand interconnect of the SGI machine. The problem points out that performing communication by means of utilizing the non-blocking pairs of MPI\_Isend() and MPI\_Irecv(), followed by MPI\_Waitall() to wait for all processes is inappropriate for the Tofu interconnect, and possibly for other interconnects that have not been examined. Therefore, sticking to a single method for handling communication is proven prone to various performance factors, and a diverse group of methods for undertaking communication should be developed for high performance and adaptability.    

\subsection{Performance Tuning}

\subsubsection{Communication Methods}

\begin{figure}[htbp]
        \centering
        \begin{subfigure}[b]{0.48\textwidth}
                \includegraphics[width=1.0\textwidth]{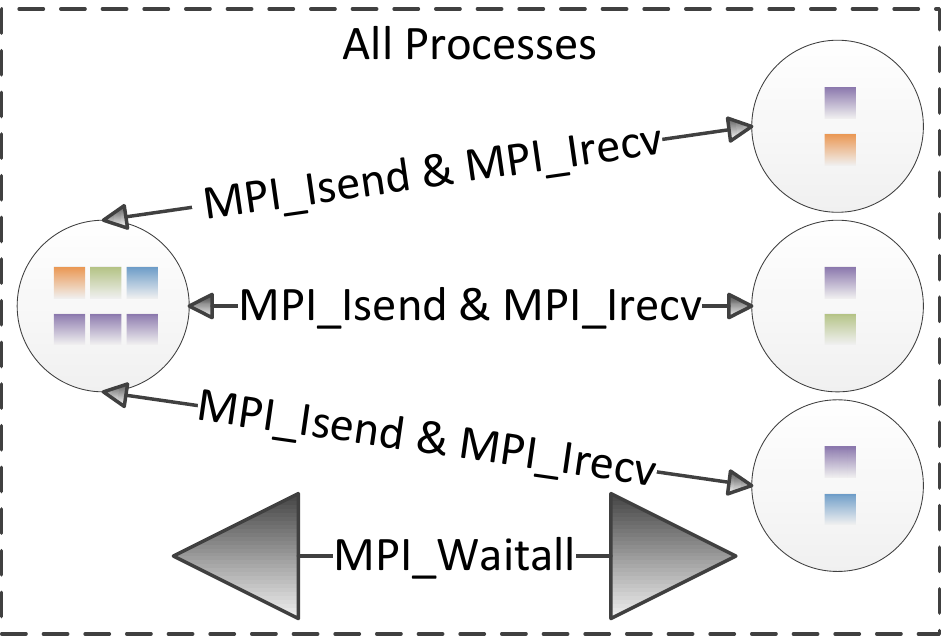}
                \caption{Isend\_Waitall.}
                \label{fig-comm-patt0}
        \end{subfigure}%
        \quad
        \begin{subfigure}[b]{0.48\textwidth}
                \includegraphics[width=1.0\textwidth]{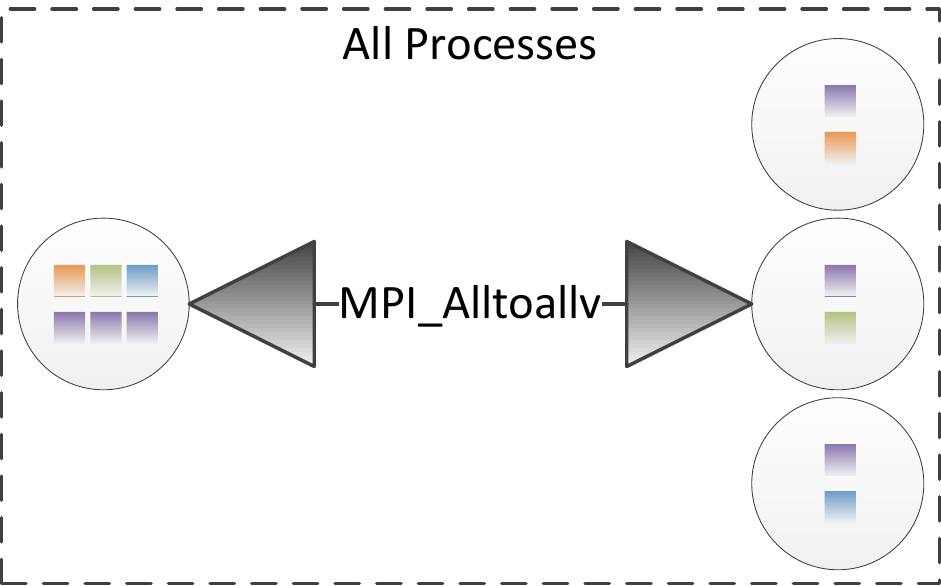}
                \caption{Alltoallv.}
                \label{fig-comm-patt1}
        \end{subfigure}%
        
        \begin{subfigure}[b]{0.48\textwidth}
                \includegraphics[width=1.0\textwidth]{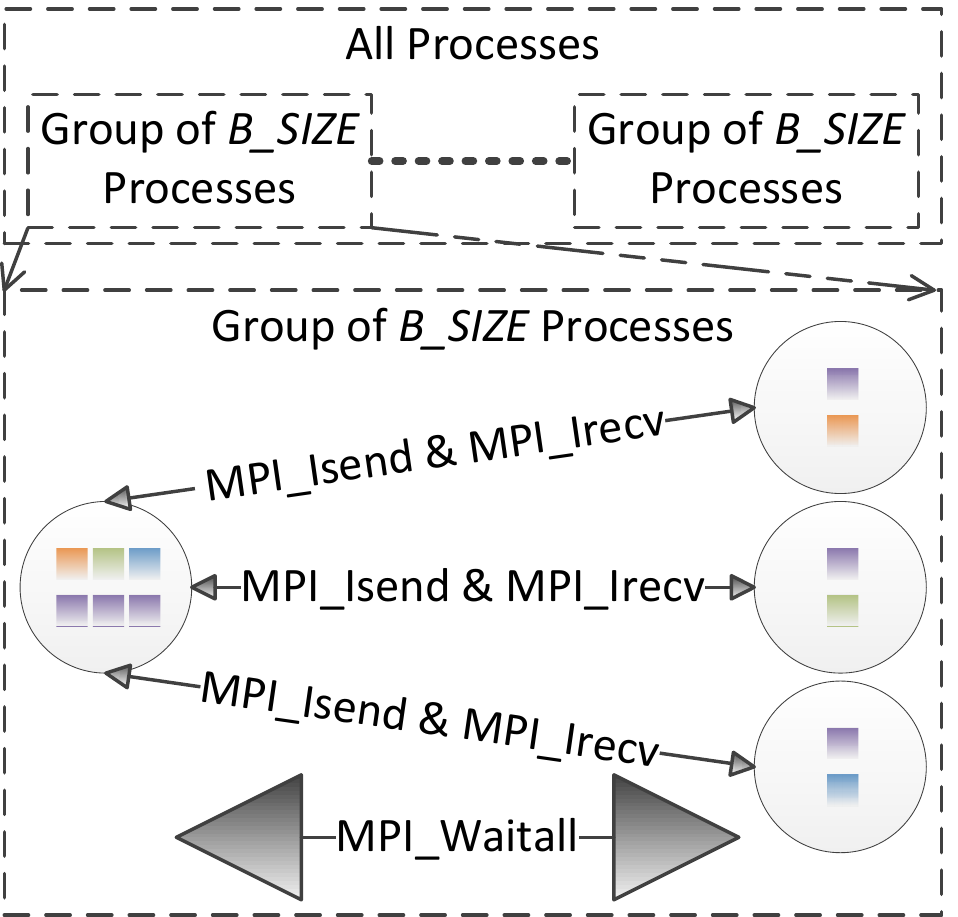}
                \caption{Isend\_Waitall\_Block.}
                \label{fig-comm-patt2}
        \end{subfigure}%
        \quad
        \begin{subfigure}[b]{0.48\textwidth}
                \includegraphics[width=1.0\textwidth]{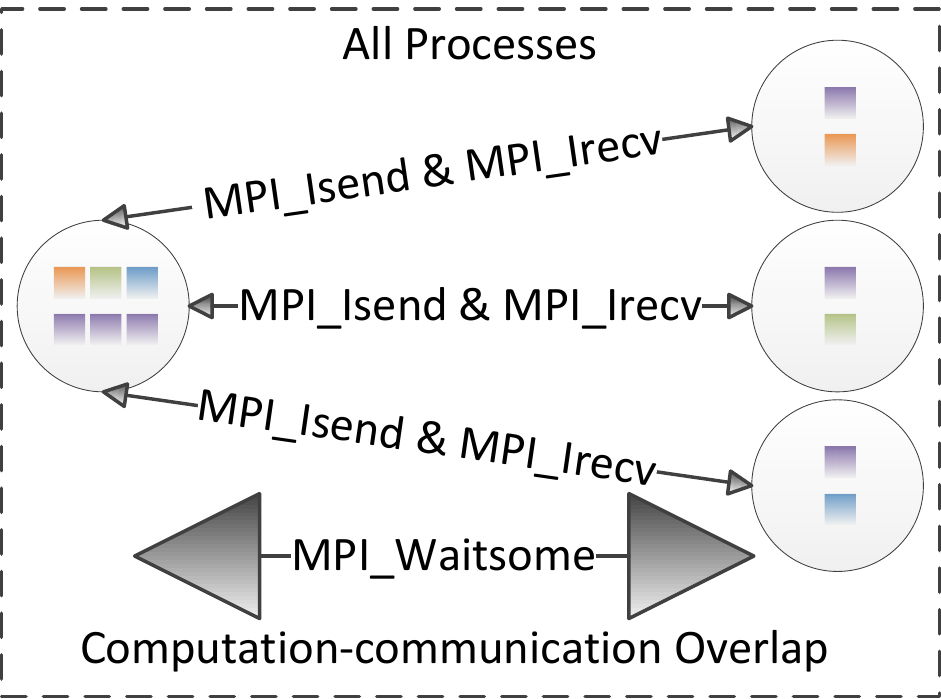}
                \caption{Isend\_Waitsome.}
                \label{fig-comm-patt3}
        \end{subfigure}%
        
        \begin{subfigure}[b]{0.48\textwidth}
                \includegraphics[width=1.0\textwidth]{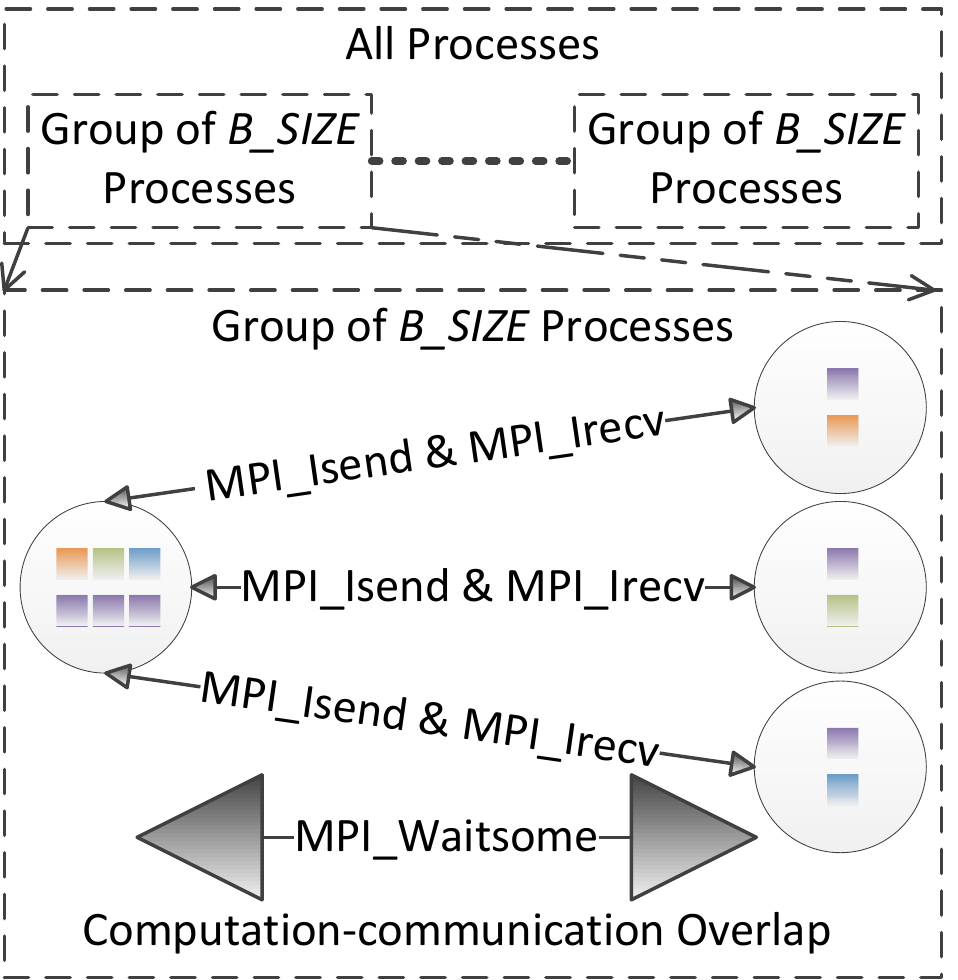}
                \caption{Isend\_Waitsome\_Block.}
                \label{fig-comm-patt4}
        \end{subfigure}%
        \quad
        \begin{subfigure}[b]{0.48\textwidth}
                \includegraphics[width=1.0\textwidth]{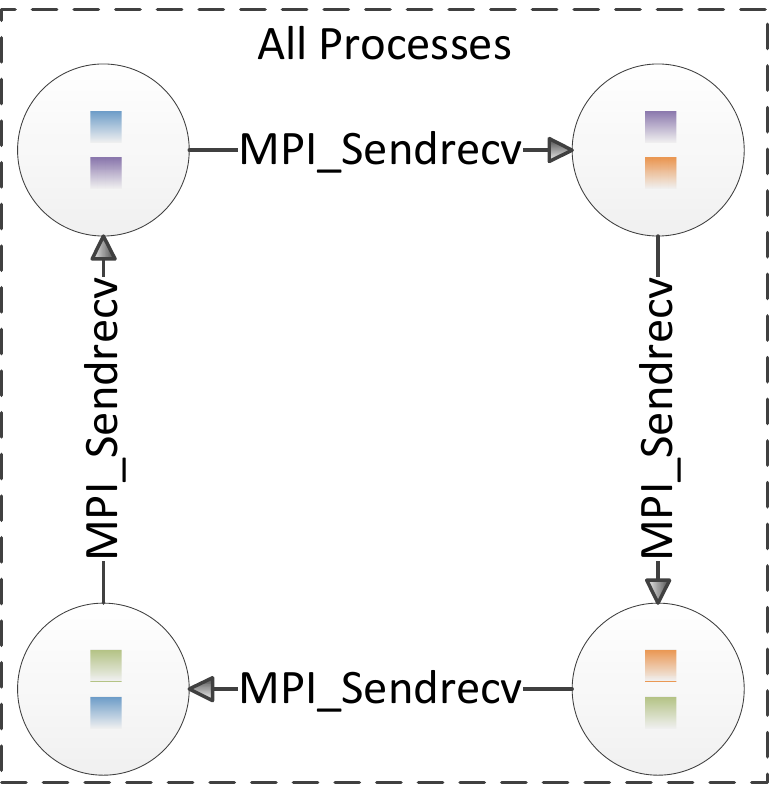}
                \caption{Sendrecv.}
                \label{fig-comm-patt5}
        \end{subfigure}%
        \caption{Communication methods.}\label{fig-comm-patt}
\end{figure}

The performance analysis suggests that performing communication by way of MPI\_Isend(), MPI\_Irecv(), and MPI\_Waitall() causes performance degradation on the FX10, and different methods for dealing with communication should be examined as a consequence.     
Figure \ref{fig-comm-patt} illustrates six communication methods that we develop for communication tuning. As there are many uncertainties in a calculation, notably the number of processes, the size of messages, and the interconnect, each method is designed to address some specific classes of calculation, for example a case with a large number of processes and a large message size. Nevertheless, although no method is expected to be the best performer for any case on any computational platform, a method that can often deliver reasonable performance may exist. Also, they are intended to cover a range of calculation scales on generic machines, rather than being optimized for a particular interconnect or machine. The methods are listed and explained below.

\begin{itemize}
\item
\textbf{Isend\_Waitall}: This is the default communication method in OpenFFT version 1.0 discussed so far, with each process calling MPI\_Isend() to send to all other processes and MPI\_Irecv() to receive from all other processes at the same time, and then MPI\_Waitall() to wait for all the processes to complete all the send and receive operations. Generally, the method is quite efficient with relatively small numbers of processes and medium message sizes. 
\item
\textbf{Alltoallv}: Communication is performed by adopting a single MPI\_Alltoallv() in all processes in a single communication step. We make use of MPI\_Alltoallv() to avoid introducing overhead on the volume of communication by padding the send and receive buffers as required by MPI\_Alltoall(). The performance of this method totally depends on the MPI implementation in the machine. Actually, we found that MPI\_Alltoall() does not perform efficiently with our decomposition method, because our method is row-based distribution and makes the volume of communication with each other process varied for a particular process. The difference can be big between the smallest and largest volumes, causing the padding to become large and expensive. 
\item
\textbf{Isend\_Waitall\_Block}: This method logically divides the processes into $N_p/B\_{SIZE}$ groups of $B\_{SIZE}$ processes that have yet to perform communications among them, where $N_p$ is the number of processes, and $B\_{SIZE}$ is a pre-defined parameter, set at 32 in our implementation. Communication is carried out within each group by having each process in the group employing MPI\_Isend() and MPI\_Irecv() to send to and receive from all other processes in the same group, and MPI\_Waitall() to wait for all the processes in the same group to complete the send and receive operations. Then the process of dividing the processes into groups and performing group communications is repeated until all communications have been finished. The number of communication steps is $N_p/B\_{SIZE}$. The method is though to be suitable for handling a very large number of processes, as it could group the communications to help prevent network congestion.   
\item
\textbf{Isend\_Waitsome}: This method also utilizes MPI\_Isend() and MPI\_Irecv() to send to and receive from all other processes, similar to Isend\_Waitall, but exploits MPI\_Waitsome() in preference to MPI\_Waitall() to enable overlapping communication with computation, which is the operation of copying data from the receive array to the local output data array (step (6) in Fig. \ref{fig-cal-flow}). As soon as a process has received data from another process, it will immediately starts unpacking the received data to the local output data array, while still receiving from other processes. In fact, the method is implemented with the aim of delivering fine performance across a spectrum of numbers of processes and message sizes due to the benefit of the overlapping.  
\item
\textbf{Isend\_Waitsome\_Block}: This method is akin to Isend\_Waitall\_Block, except that MPI\_Waitsome() is applied rather than MPI\_Waitall() for enabling communication-computation overlap like Isend\_Waitsome. Communication is carried out within each group by having each process in the group employing MPI\_Isend() and MPI\_Irecv() to send to and receive from all other processes in the same group. By utilizing MPI\_Waitsome(), a process can immediately unpack the data received from another process to the local output data array once it has arrived. The method augments Isend\_Waitall\_Block by taking advantage of the overlapping, and may effectively address communication with a large number of processes and long messages. 
\item
\textbf{Sendrecv}: In this method, only point-to-point communication is conducted between pairs of processes using MPI\_Sendrecv(). At step $i$th, the process with the process identification $myid$ sends a message to the process with the process identification $myid+i$, and receives
a message from the process with the process identification $myid-i$, with wrap around for a total of $N_p$ communication steps. It is expected to give good performance when long messages are transferred over the interconnection network.  
\end{itemize}

It is worth noting that these communication methods are far from exhaustive. There are other methods, for example those with the non-blocking collective and one-sided communication routines that are designed to improve communication. However, we must also weigh the importance of portability, and decide to leave them in future work, when they are readily available in all popular MPI libraries.   

\subsubsection{Performance of Communication Methods}
\label{sec-per-comm}

\begin{figure}[htbp]
        \begin{subfigure}{\textwidth}
                \centering
                \includegraphics[width=0.75\textwidth]{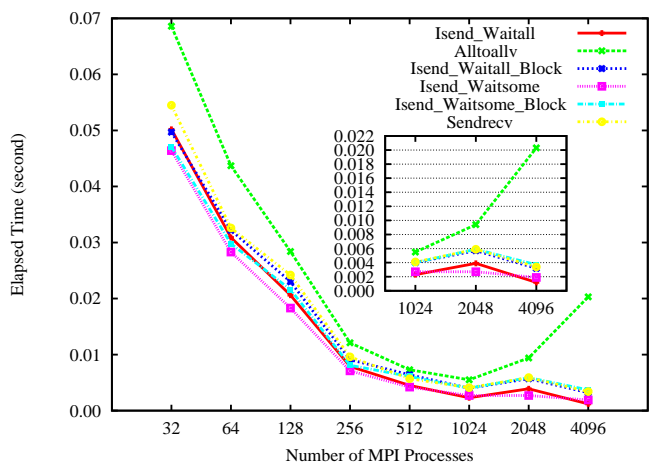}
                \caption{Cray XC30. The inset enlarges the right part of the main graph.}
                \label{fig-tuning-cray}
        \end{subfigure}%
        
        \begin{subfigure}{\textwidth}
                \centering
                \includegraphics[width=0.75\textwidth]{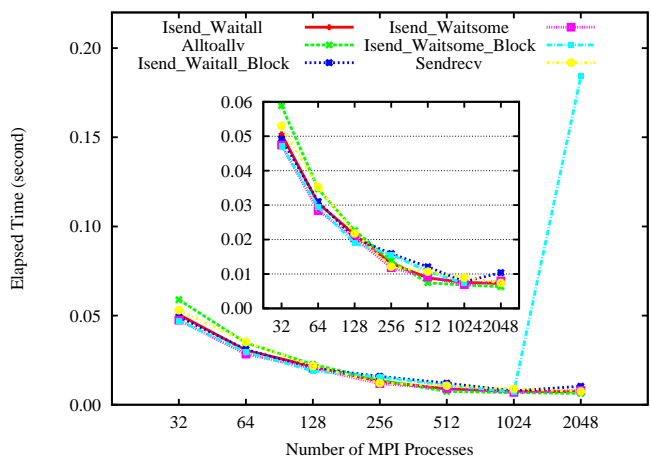}
                \caption{SGI InfiniBand. The inset enlarges the main part with some extreme points removed.}
                \label{fig-tuning-hster}
        \end{subfigure}

        \begin{subfigure}{\textwidth}
                \centering
                \includegraphics[width=0.75\textwidth]{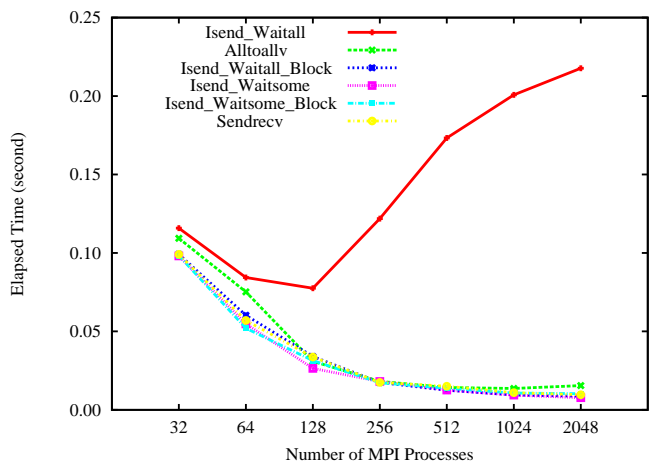}
                \caption{Fujitsu FX10.}
                \label{fig-tuning-fx10}
        \end{subfigure}%
        \caption{Tuning of communication with $256^3$ data points on the Cray XC30, SGI InfiniBand, and Fujitsu FX10.}\label{fig-tuning-comm}
\end{figure}

We implement the six communication methods described above in OpenFFT, and perform double-precision complex-to-complex transforms with the representative size of $256^3$ data points to compare their performance on the machines. 
Figure \ref{fig-tuning-comm} shows the performance comparison on the Cray XC30 (Fig. \ref{fig-tuning-cray}), SGI InfiniBand (Fig. \ref{fig-tuning-hster}), and Fujitsu FX10 (Fig. \ref{fig-tuning-fx10}). We make the following observations. 

\begin{itemize}
\item
\textbf{Certain methods are ineffective and should not be used on a particular machine}. Figure \ref{fig-tuning-cray} shows that Alltoallv is always the worst method on the XC30, implying that MPI\_Alltoallv() in the MPI implementation of the machine may not be as heavily optimized as those in the other two machines for a large number of processes. On the SGI InfiniBand (Fig. \ref{fig-tuning-hster}), the method Isend\_Waitsome\_Block must be utilized carefully, as it could unexpectedly lead to a sudden large drop in performance when a lot of processes are used. The outcome is caused by the combination of the group size of 32 processes and overlapping, which is too small to take advantage of overlapping and shown to generate overhead of the grouping to the InfiniBand interconnect. On the other hand, Isend\_Waitall is unsuited to the FX10, as pointed out earlier, and Alltoallv should also be avoided if possible. The utilization of MPI\_Waitall to wait for all processes in Isend\_Waitall may cause unbalanced wait times among the processes in the FX10, where some finish much earlier than the others, that grow to the number of processes, leading to larger performance deterioration with more processes.  
\item
\textbf{Certain methods are good on a machine, yet turn out to be bad on another}. Isend\_Waitall generally delivers fine performance on the XC30 and SGI InfiniBand, but at the same time is the worst method on the FX10. This indicates that the XC30 and SGI InfiniBand machines do not experience unbalanced wait times as the FX10 does. Likewise, Alltoallv is rather efficient on the SGI InfiniBand, especially with a large number of processes, while being the worst and second worst on the XC30 and FX10, respectively. As Alltoallv totally depends on the MPI implementation for performance, the MPI implementation of the SGI InfiniBand is considered to be more optimized than its counterparts of other machines. The result confirms the impact of machine specifications, including the interconnect and system software, in applications' performance.
\item
\textbf{Isend\_Waitsome appears to be the most stable performer}. Although there is no always-best method, Isend\_Waitsome demonstrates its stability and can generally give good results at different calculation scales on the machines, thanks to the ability of exploiting the communication-computation overlap, unlike Isend\_Waitsome\_Block that suffers from the process grouping overhead with many processes.  

\end{itemize}

\subsubsection{Auto-Tuning of Communication}

The comparison results of the communication methods again assert that adhering to one specific method is sensitive to unexpected performance degradation. Also, even though Isend\_Waitsome is shown to be stable, there are still cases when other methods have the edge over it. That said, a fair selection, where all the methods are taken into account, is desirable if possible. Unfortunately, sophisticated model-driven and heuristic-based optimization methods, such as the one in  \cite{Pouchet:2010:CIM:1884643.1884672}, are suitable only for single-node tuning techniques with compilers, for instance loop transformation and thread parallelization, rather than multi-node communication optimization. This leads us to develop an auto-tuning feature in OpenFFT. 
When the feature is enabled, we will perform the calculation a few times with the six communication methods, and then select the best performer during the initialization phase of the calculation in run time. If the auto-tuning of communication is disabled, the default communication method, which is Isend\_Waitsome, will be chosen. To further provide flexibility, a User-select option is added to the auto-tuning feature to allow one to select any of the six methods. Differently from the tuning of collective operations \cite{Thakur03improvingthe} that considers the sole metric of the message size to choose the algorithm, in doing so, our auto-tuning feature takes into account the number of processes participating in the communication and the communication-computation overlapping, in addition to the size of messages.   

To achieve the highest possible performance, it is recommended to enable the auto-tuning, in exchange for overhead in the initialization phase. The overhead is thought negligible, though, as the number of executions for the run-time selection of communication method is usually far smaller than that of calling the execution phase in practical scenarios. In particular, there are six methods with each executed twice, for instance, in the auto-tuning process, resulting in $6 \times 2=12$ times of execution, as against the norm of tens to hundreds of times for carrying out the execution phase. For example, assume that the execution phase is conducted for 100 times, then the overhead of auto-tuning accounts for $12/(12+100)=10.71$\% of the total execution time. Furthermore, the overhead of auto-tuning can be minimized by first enabling the feature to obtain the best method with a specific machine and problem size, and then utilizing the User-select option to always specify and use that method thereupon. In this case, the overhead of auto-tuning disappears. 

With the auto-tuning of communication, we aim to cover a spectrum of calculation scales on different machines to make it possible for OpenFFT to maintain relatively good performance, even with machines it has never been investigated.

\subsubsection{Optimization of Array Allocations}

In addition to the tuning of communication, we also optimize the allocation and de-allocation of the temporary arrays in the execution phase. In OpenFFT version 1.0, the temporary arrays for the FFTW and communication buffers, as well as the arrays for MPI requests and statues, were repeatedly allocated and de-allocated during the execution phase. This practice is viewed as potentially harmful to the performance. In version 1.1, we reduce the number of temporary arrays by exploiting global common arrays, and move their allocation and de-allocation to the initialization phase. In doing so, we need to allocate and de-allocate them only once, before and after the calculation.  

\section{Results}
In this section, we undertake performance benchmarks with OpenFFT and several state-of-the-art 2-D packages on the machines. The following official versions are utilized at the time of writing.
\begin{itemize}
\item
\textbf{OpenFFT}: version 1.1 at http://www.openmx-square.org/openfft/, with auto-tuning of communication enabled. 
\item
\textbf{2DECOMP\&FFT}: version 1.5.847 at http://www.2decomp.org/, with auto-tuning of decomposition enabled. 
\item
\textbf{P3DFFT}: version 2.7.1 at https://code.google.com/p/p3dfft/. As there is no support for complex-to-complex transforms, the r2c interface is used with 2x real numbers for the equivalent of 1x complex numbers. 
\item
\textbf{FFTE}: version 6.0 at http://www.ffte.jp/. The 2-D decomposition version (pzfft3dv) is adopted, together with its own FFT engine and process grid. 

\end{itemize}

In the benchmarks, double-precision complex-to-complex transforms are performed with the same version of FFTW \cite{FFTW} as the 1-D FFT engine (refer to table \ref{tab-spec-machine} for the version of FFTW on each machine), except for P3DFFT, where the r2c interface is fed with 2x real numbers, and FFTE, which employs its own FFT engine. With OpenFFT, the auto-tuning of communication is enabled to opt for the best one among the six communication methods in run time for each combination of the number of processes and data size. Similarly, the auto-tuning of decomposition of 2DECOMP\&FFT is adopted to obtain the optimally estimated process grid for each combination, which is then applied in the calculations in 2DECOMP\&FFT and P3DFFT. We also separately perform 2DECOMP\&FFT and P3DFFT with the 1-D decomposition (1-D), where the process grid is $1 \times N_p$. 

\subsection{Volume of Communication}

\begin{figure}[b]
\begin{center}
\includegraphics[scale=0.95]{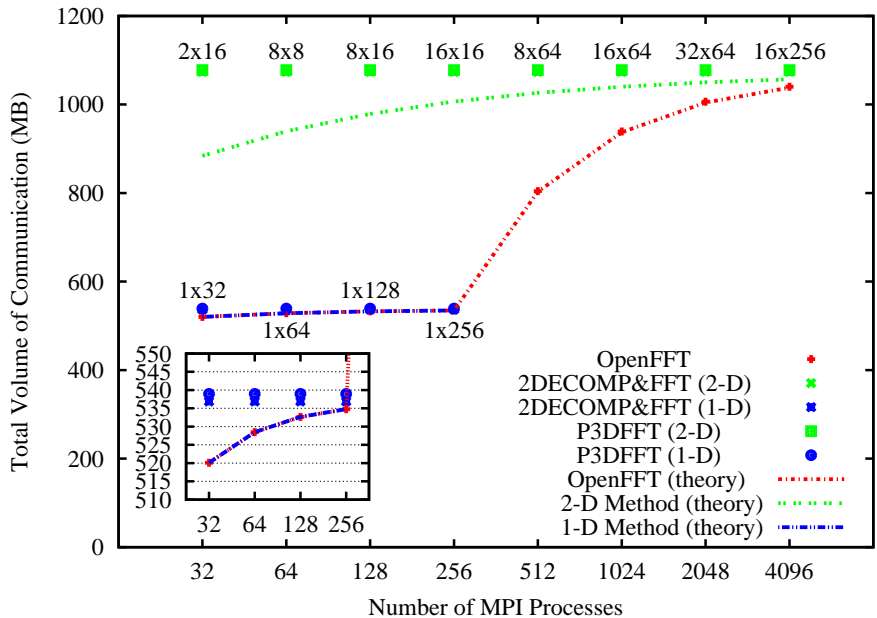}
\end{center}
\caption{Total volume of communication: comparison between OpenFFT and other packages in theory and practice with $256^3$ data points in double-precision complex-to-complex transforms.}
\label{fig-compare-volume}
\end{figure}

To complement our previous work \cite{Duy2014153} that did not show the measured values in practice, let us start with presenting the total volume of communication of all the processes in both theory and practice incurred by OpenFFT and two other representative packages in Fig. \ref{fig-compare-volume}. The theoretical volumes are given by \cite{Duy2014153}, multiplied by a factor of ($N_p \times 16 \times 2$), because they are for all the processes with complex-to-complex transforms, and double in size as the summation of equal send and receive volumes measured by MPI profilers. The practical volumes are collected by the MPI profiler on the K computer, and confirmed by the profiler on the Cray XC30 with $256^3$ data points. The volumes of 2DECOMP\&FFT and P3DFFT are recorded in two separate cases of decomposition: 2-D decomposition (2-D) and 1-D decomposition (1-D). The text labels along the line points indicate the process grid, for instance 2$\times$16 means there are 32 processes arranged in a grid of 2 rows and 16 columns. On the other hand, the volumes of OpenFFT are taken with the communication method Isend\_Waitall. Other communication methods should have similar volumes, except for Isend\_Waitsome and Isend\_Waitsome\_Block that may incur some extra, but tiny and negligible, messages caused by the use of MPI\_Waitsome() for polling the receive buffer for message arrival. Again, the size of $256^3$ is chosen as a representative case, and calculations for other sizes' volumes are straightforward. 

It is obvious in Fig. \ref{fig-compare-volume} that OpenFFT is always more communication-efficient than the other packages in practice, despite the decomposition of 1-D or 2-D. The difference in the volumes gradually becomes smaller and smaller with an increase in the number of processes. The figure also demonstrates that the decomposition method of OpenFFT is adaptive with the 1-D decomposition for up to 256 processes and the 2-D for the rest. There is a surge in its volume from 256 to 512 processes, due to the switching from 1-D to 2-D of the decomposition. In addition, the theoretical and practical volumes of OpenFFT exactly match each other, as a result of the design and development strategy of a communication-aware package, while the others incur slightly higher volumes in practice than the theoretical volumes owing to the padding of the communication arrays.

\subsection{Numerical Comparison}

\begin{figure}[htb]
\begin{center}
\includegraphics[scale=0.7,trim = 13mm 20mm 0mm 0mm]{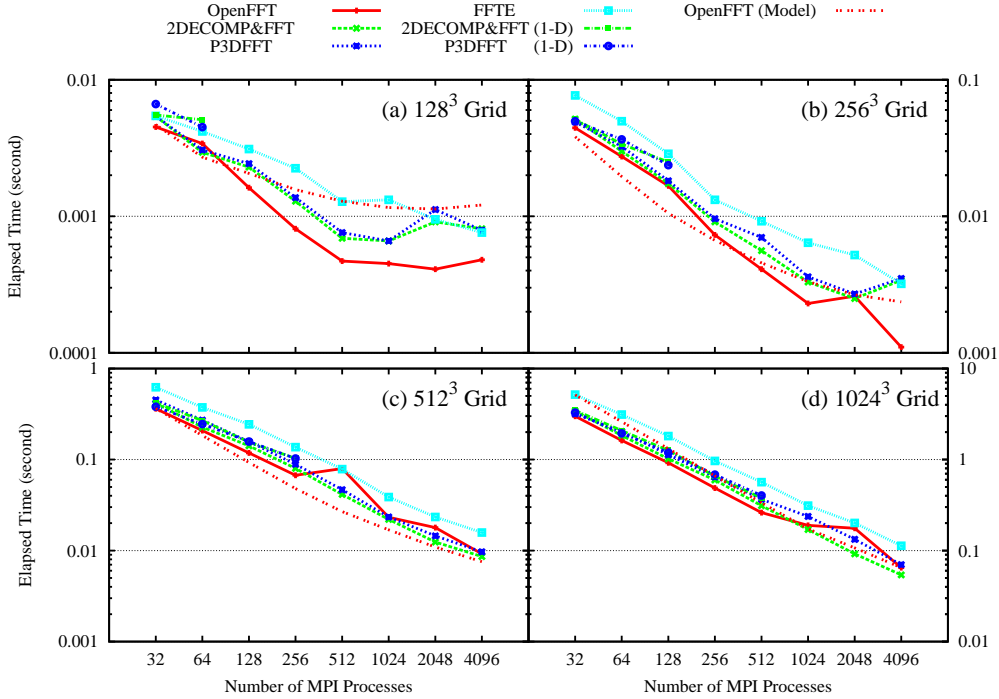}
\end{center}
\caption{Numerical comparison on the Cray XC30.}
\label{fig-compare-cray}
\end{figure}

\begin{figure}[htb]
\begin{center}
\includegraphics[scale=0.7,trim = 13mm 20mm 0mm 0mm]{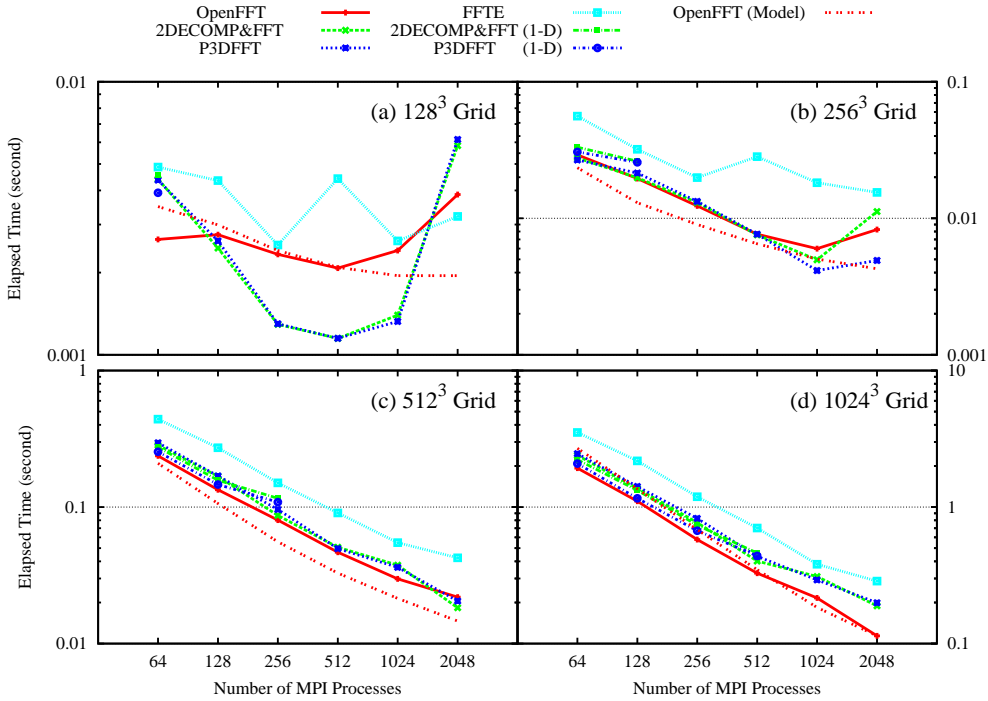}
\end{center}
\caption{Numerical comparison on the SGI InfiniBand.}
\label{fig-compare-hster}
\end{figure}

\begin{figure}[htb]
\begin{center}
\includegraphics[scale=0.7,trim = 13mm 20mm 0mm 0mm]{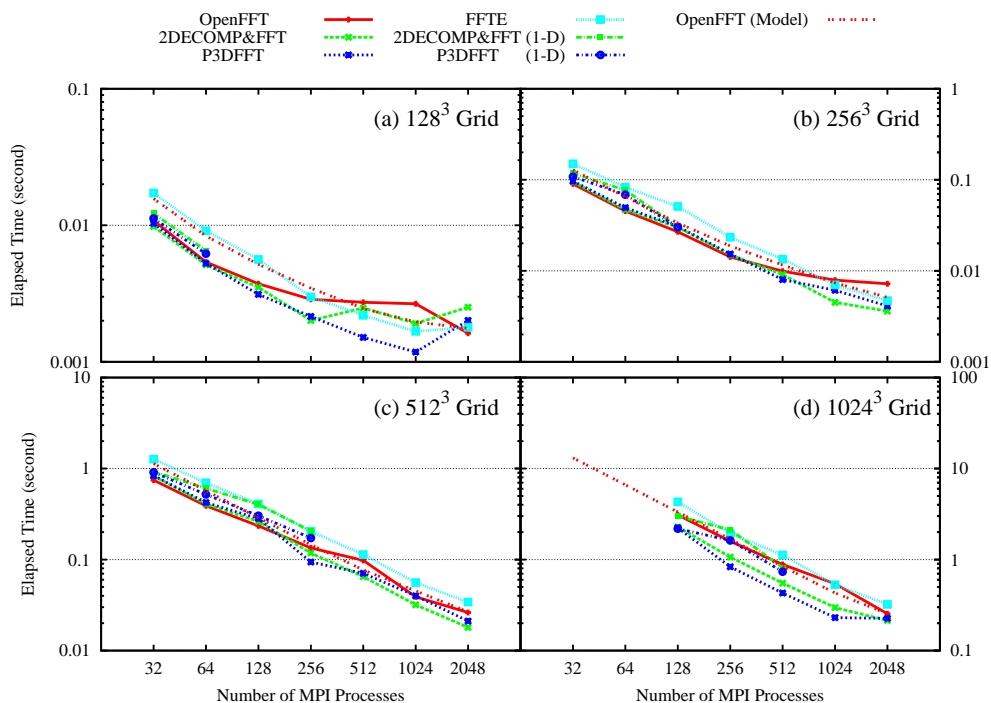}
\end{center}
\caption{Numerical comparison on the Fujitsu FX10.}
\label{fig-compare-fx10}
\end{figure}

\begin{figure}[htb]
\begin{center}
\includegraphics[scale=0.7,trim = 13mm 20mm 0mm 60mm]{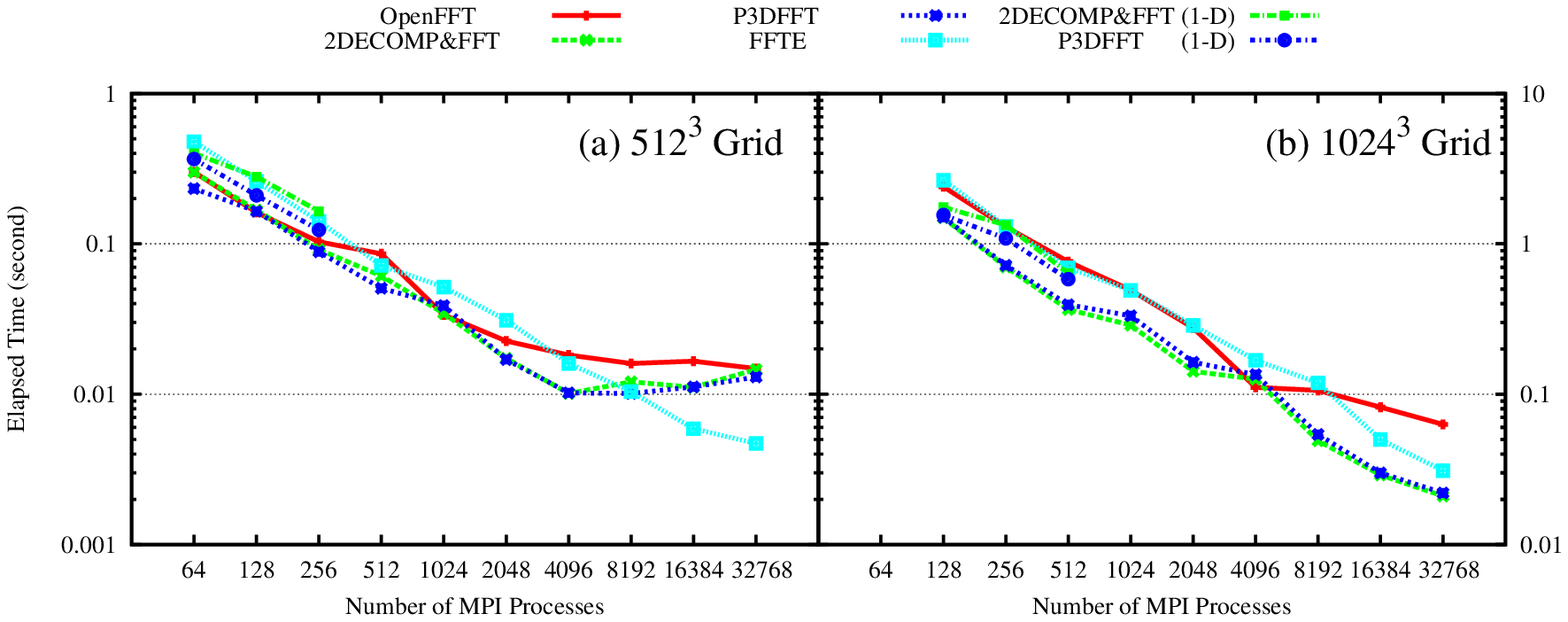}
\end{center}
\caption{Numerical comparison on the K computer.}
\label{fig-compare-k}
\end{figure}

Figures \ref{fig-compare-cray}, \ref{fig-compare-hster}, \ref{fig-compare-fx10}, and \ref{fig-compare-k} display the time-to-solution performance of the packages on the Cray XC30, SGI InfiniBand, Fujitsu FX10, and K computer, respectively. In the benchmarks, the main FFT execution routines of the packages are performed for ten times, and the average of the ten longest process times is reported, excluding the initialization and finalization times. Moreover, each set of benchmark is repeated in multiple times to obtain the most stable results in the flat MPI mode by placing one MPI process on one core. The packages are evaluated from small-to-medium sizes of $128^3$ and $256^3$ to medium-to-large sizes of $512^3$ to $1024^3$. 
We note some important results on the machines as follows.

\begin{itemize}
\item
\textbf{Cray XC30}. Figures \ref{fig-compare-cray}a and \ref{fig-compare-cray}b show that OpenFFT outperforms other packages at almost all calculation scales for the smaller data sizes of $128^3$ and $256^3$. The performance difference is quite noticeable for the smallest size of $128^3$. A likely explanation is that while the other codes mainly use MPI\_Alltoall and MPI\_Alltoallv, OpenFFT implements a wider range of communication methods. As discussed in section \ref{sec-per-comm}, MPI\_Alltoallv on the XC30 does not perform as well as other methods at this scale of calculation, and this may be the reason behind the difference. It can also be credited to the impact and advantage of having smaller volumes of communication on the Dragonfly interconnect, since the smaller the data size is, the larger the impact of communication is. With the medium-to-large sizes of $512^3$ and $1024^3$, however, the performance of the packages appears to be quite comparable with little performance difference, when the impact of communication becomes smaller with an increase in data size. In addition, OpenFFT experiences some performance drops in the cases of $512^3$ with 512 processes and $1024^3$ with 2,048 processes, caused by a surge in the communication time, perhaps due to a sudden overhead on the interconnect.     
\item
\textbf{SGI InfiniBand}. The results on this machine can be summarized in the following three observations: OpenFFT is worse than other packages with the smallest size of $128^3$ (Fig. \ref{fig-compare-hster}a), is almost similar to and worse than with the size of $256^3$ (Fig. \ref{fig-compare-hster}b), and is slightly superior for most cases with the sizes of $512^3$ and $1024^3$ (Figs. \ref{fig-compare-hster}c and d). 2DECOMP\&FFT and P3DFFT are better with $128^3$, which is opposite to that on the Cray XC30, that can be interpreted from the high performance of MPI\_Alltoall() of the SGI InfiniBand machine applied in the codes, as also touched on in section \ref{sec-per-comm}. The results with $512^3$ and $1024^3$, which has a smaller communication impact, look very similar to those on the Cray XC30. In terms of scalability, all the codes initially suffer performance degradation when using more than 1,000 processes with $128^3$, which proves to be too small on the machine, but could scale efficiently with the larger sizes, especially with $512^3$ and $1024^3$, possibly driven by the InfiniBand interconnect.   
\item
\textbf{Fujitsu FX10}. Compared with the Cray XC30 and SGI InfiniBand (table \ref{tab-spec-machine}), the FX10 has a lower CPU performance (1.848GHz as against 2.6 and 2.8GHz) and a narrower announced bandwidth of the interconnect (5.0GB/s as against 8.5 and 6.8GB/s). Hence, the total elapsed times of the packages on the FX10 are always longer than their corresponding counterparts obtained on the Cray XC30 and SGI InfiniBand. In the cases of $128^3$ (Fig. \ref{fig-compare-fx10}a), $256^3$ (Fig. \ref{fig-compare-fx10}b), and $512^3$ (Fig. \ref{fig-compare-fx10}c), OpenFFT usually starts relatively effectively with smaller numbers of processes, but does not maintain the performance for long and deteriorates when more processes are employed. The case of $1024^3$ (Fig. \ref{fig-compare-fx10}d) seems to be different, where it appears to begin improving with more than 2000 processes, where its scalibility becomes slightly better.   
\item
\textbf{K computer}. We aim to perform large scale calculations on the K computer, and thus, using up to 32,768 processes for the medium-to-large sizes of $512^3$ (Fig. \ref{fig-compare-k}a) and $1024^3$ (Fig. \ref{fig-compare-k}b). As the FX10 can actually be viewed as a smaller sibling of the K computer with many identical specifications, the behaviors of the packages on the K computer for up to 2,048 processes are observed to closely resemble those on the FX10. From 4,096 processes, although the packages, except for FFTE, hardly scale well with $512^3$, their scalalibity is quite good with $1024^3$, thanks to a higher computation to communication ratio. Also, one can see that FFTE achieves its best performance on the K computer, likely as a result of effective utilization of the configurable virtual 3-D torus topology network from users' programming point of view, which is a distinguishing feature of the machine. OpenFFT has longer elapsed time than those of other packages from the beginning with the case of $1024^3$, different from the results acquired on the Cray XC30 and SGI InfiniBand. Since OpenFFT produces a string of cache misses for packing data to the send buffer and for unpacking data from the receive buffer, its performance is affected by the cache miss penalty, which seems to be higher to a larger data size on the K computer and FX10 than that on the Cray XC30 and SGI InfiniBand. 
\end{itemize}

In summary, OpenFFT is able to deliver relatively good results on the Cray XC30 and SGI InfiniBand, which are popular Intel-based machines, with the InfiniBand machine basically belonging to the class of universal general-purpose Linux clusters. Although its performance is not as comparatively satisfactory on the Fujitsu-made machines, i.e. the FX10 and K computer, OpenFFT is still thought to give a fairly reasonable scalability. The overall performance can be attributed to the design and implementation of our communication-optimal method. 

\section{Conclusion}
\label{Conclusion}
In this paper, we have analyzed and tuned the performance of OpenFFT, an open-source parallel package for 3-D FFTs. Given the performance analysis, six communication methods that are designed to cover a range of calculation scales on different computational platforms in performing communication have been developed. We have augmented OpenFFT with the auto-tuning of communication, where the best method is chosen in run time based on their performance. The optimized OpenFFT, released as OpenFFT version 1.1, has been shown to be capable of achieving fairly good performance at small-to-large computational scales on a diverse group of machines. The communication methods are by no means limited to OpenFFT, and can be applied in general parallel applications.   
In future work, we plan to implement the newer non-blocking collective and one-sided communication routines, and extend OpenFFT to provide support for higher-than-3D parallel FFTs with different kinds of transform. 



\begin{acknowledgements}
This work was supported by the Strategic Programs for Innovative Research (SPIRE), MEXT, the Computational Materials Science Initiative (CMSI), and Materials Design through Computics: Complex Correlation and Non-Equilibrium Dynamics A Grant in Aid for Scientific Research on Innovative Areas, MEXT, Japan. 
The benchmark calculations were performed using the K computer at RIKEN, the Fujitsu FX10 at The University of Tokyo, and the Cray XC30 and SGI InfiniBand machines at Japan Advanced Institute of Science and Technology (JAIST). We also thank Prof. Katsumi Hagita of National Defense Academy of Japan for helpful discussions. 
\end{acknowledgements}



\bibliographystyle{spmpsci}
\bibliography{biblio}







\end{document}